\documentclass[sigconf]{acmart}
\usepackage{multirow}
\usepackage{xcolor}
\usepackage{dutchcal}
\usepackage{amsfonts}
\usepackage{amsmath}
\usepackage{algorithm}
\usepackage{algpseudocode}
\usepackage{balance}

\def\Data{\mathcal{D}}
\def\Bfx#1{\boldsymbol{x}_{#1}}
\def\Bf#1{\boldsymbol{#1}}
\DeclareMathOperator*{\argmax}{arg\,max}
\DeclareMathOperator*{\onehot}{one\,hot}

\AtBeginDocument{%
  \providecommand\BibTeX{{%
    \normalfont B\kern-0.5em{\scshape i\kern-0.25em b}\kern-0.8em\TeX}}}

\copyrightyear{2023}
\acmYear{2023}
\setcopyright{acmlicensed}\acmConference[CIKM '23]{Proceedings of the 32nd ACM International Conference on Information and Knowledge Management}{October 21--25, 2023}{Birmingham, United Kingdom}
\acmBooktitle{Proceedings of the 32nd ACM International Conference on Information and Knowledge Management (CIKM '23), October 21--25, 2023, Birmingham, United Kingdom}
\acmPrice{15.00}
\acmDOI{10.1145/3583780.3614925}
\acmISBN{979-8-4007-0124-5/23/10}

\begin{document}

\title[iHAS: Instance-wise Hierarchical Architecture Search for DLRMs]{iHAS: Instance-wise Hierarchical Architecture Search \\ for Deep Learning Recommendation Models}

\author{Yakun Yu}
\email{yakun2@ualberta.ca}
\affiliation{%
  \institution{University of Alberta}
  \city{Edmonton}
  \state{Alberta}
  \country{Canada}}

\author{Shi-ang Qi}
\email{shiang@ualberta.ca}
\affiliation{%
  \institution{University of Alberta}
  \city{Edmonton}
  \state{Alberta}
  \country{Canada}}

\author{Jiuding Yang}
\email{jiuding@ualberta.ca}
\affiliation{%
  \institution{University of Alberta}
  \city{Edmonton}
  \state{Alberta}
  \country{Canada}}

\author{Liyao Jiang}
\email{liyao1@ualberta.ca}
\affiliation{%
  \institution{University of Alberta}
  \city{Edmonton}
  \state{Alberta}
  \country{Canada}}

\author{Di Niu}
\email{dniu@ualberta.ca}
\affiliation{%
  \institution{University of Alberta}
  \city{Edmonton}
  \state{Alberta}
  \country{Canada}}

\renewcommand{\shortauthors}{Yakun Yu, Shi-ang Qi, Jiuding Yang, Liyao Jiang, \& Di Niu}

\begin{abstract}
  Current recommender systems employ large-sized embedding tables with uniform dimensions for all features, leading to overfitting, high computational cost, and suboptimal generalizing performance. 
  Many techniques aim to solve this issue by feature selection or embedding dimension search. 
  However, these techniques typically select a fixed subset of features or embedding dimensions for all instances and feed all instances into one recommender model without considering heterogeneity between items or users.
  This paper proposes a novel instance-wise Hierarchical Architecture Search framework, iHAS, which automates neural architecture search at the instance level.
  Specifically, iHAS incorporates three stages: searching, clustering, and retraining. 
  The searching stage identifies optimal instance-wise embedding dimensions across different field features via carefully designed Bernoulli gates with stochastic selection and regularizers. 
  After obtaining these dimensions, the clustering stage divides samples into distinct groups via a deterministic selection approach of Bernoulli gates.  
  The retraining stage then constructs different recommender models, each one designed with optimal dimensions for the corresponding group. 
  We conduct extensive experiments to evaluate the proposed iHAS on two public benchmark datasets from a real-world recommender system. The experimental results demonstrate the effectiveness of iHAS and its outstanding transferability to widely-used deep recommendation models.
\end{abstract}





\maketitle

\section{Introduction}
Recommender systems, which aim to predict the preference of users, have been widely deployed in various real-world scenarios, e.g., online advertising~\cite{richardson2007predicting, chapelle2014simple},
social media~\cite{guy2010social}, 
news apps~\cite{zheng2018drn}, etc. 
Deep learning recommendation models (DLRMs) typically take a large amount of categorical (e.g., gender) or numerical (e.g., age) field features as input. 
These features are first encoded into high-dimensional sparse one-hot vectors, which are later mapped into real-valued dense vectors via embedding tables.
The recommender model then feeds these embeddings into a feature interaction layer which usually consists of deep neural network (DNN) \cite{covington2016deep} or factorization machine (FM) \cite{guo2017deepfm, lian2018xdeepfm, xiao2017attentional, rendle2010factorization} to model user preferences for final prediction.

The embedding tables play a fundamental role in the recommendation system, as they dominate the majority of parameters. 
However, most existing methods construct their proposed recommender models with large-sized embedding tables and a uniform dimension size for all possible fields~\cite{guo2017deepfm, he2017neural, song2019autoint}, which may lead to overfitting, high computational cost, and poor model generalization \cite{wang2022autofield, lin2022adafs, qu2022single}.
Therefore, the first objective for an optimal DLRM is to find optimal embedding dimensions for different fields and remove redundant dimensions. 
Performing embedding dimension search is also sufficient to include feature selection, i.e., the optimal dimension for a field feature could be zero, which means completely excluding this feature.

A common approach for embedding dimension search is to employ the $\mathcal{l}_0$ norm on the dimensions to penalize the count of non-zero dimension entries. 
However, as the $\mathcal{l}_0$ norm poses a computational challenge for gradient descent,
researchers have attempted to substitute $\mathcal{l}_0$ with a surrogate function, such as the $\mathcal{l}_1$ norm for LASSO~\cite{tibshirani1996regression}, yet achieving limited selection ability~\cite{yamada2020feature}.
Recently, probabilistic approaches~\cite{louizoslearning, wang2022autofield} suggest utilizing Bernoulli random variables (RVs) with Gumbel-Softmax approximations to identify the top $K$ features with the highest probabilities. 
Though these methods can be applied for dimension selection, we have empirically observed that the probabilities learned are often indistinguishable. Therefore, selecting the top $K$ features/dimensions based on these probabilities may inadvertently result in either the exclusion of critical features/dimensions or the inclusion of irrelevant ones.

Furthermore, prior approaches uniformly apply embedding dimension selection across all instances in the datasets, therefore disregarding the inherent variations between individual samples. 
This one-size-for-all approach can be inadequate in many scenarios, especially when dealing with highly heterogeneous populations where relevant features can significantly diverge across users or across items.
For example, in a movie recommendation system, the feature ``age'' usually plays a crucial role in recommending Disney movies, thereby possibly necessitating a larger embedding dimension. Conversely, ``age'' is less relevant for comedy films, resulting in a smaller dimension size.
Thus, it is evident that treating all instances identically may overlook these context-specific nuances.
Intuitively speaking, when dimension selection is performed at the instance level, we can create neural architectures that are better suited to individual samples. 
Such an approach not only results in superior performance but also enables faster inference times by focusing on the most relevant dimensions of each sample.

In this paper, we propose an instance-wise Hierarchical Architecture Search framework, iHAS, which attempts to perform automatic architecture search on the instance level, using hierarchical training procedures for DLRMs. 
Specifically, iHAS includes three learning stages: searching, clustering, and retraining. 
The searching stage aims to find the optimal instance-wise embedding dimensions across different fields via a carefully designed ``Bernoulli gate'' with stochastic selection mode and a regularizer. 
After selecting instance-wise embedding dimensions, we separate samples into different groups based on a novel deterministic selection approach in the clustering stage. 
The retraining stage trains different recommender models, with optimal dimensions tailored to different groups. 
During inference time, each test sample will first be assigned to a suitable group, where predictions are made by the corresponding recommender model. 
We summarize our major contributions as:
\begin{itemize}
    \item We propose a hierarchical training framework that uses instance-wise ``Bernoulli gates'' to facilitate effective dimension search for each sample.
    \item We apply a sparse and bi-polarization regularizer in the objective function to help the model learn distinguishable Bernoulli RVs, and use a threshold selector for downstream deterministic selection.
    \item To balance the trade-off between the one-size-for-all and full-customization (which is not applicable with finite data size) strategies, we propose to divide samples into clusters and develop tailored recommender models for each cluster.
\end{itemize}
We empirically evaluate the performance of our framework on two large-scale benchmark datasets. 
Our experimental results indicate a notable superiority of our approach over various state-of-the-art baseline models on both datasets.
Furthermore, the transferability analysis demonstrates our framework can be effectively transferred to diverse deep recommender models, thereby enhancing their performance.
Additionally, our framework 
offers an efficiency advantage as it requires less inference time than competing baseline models.

\section{Related Work}
This section introduces the main related works to our study, focusing on feature-based recommender models and AutoML approaches for recommendation systems. 

\subsection{Feature-based Recommender Models}
Feature-based recommender models take sparse, high-dimensional features from users and items as input and transform them into low-dimensional representations to capture user preferences for improved recommendations.
For example, Cheng et al.~\cite{cheng2016wide} propose Wide\&Deep (W\&D), a model composed of a linear module and a Multi-Layer Perceptron (MLP) layer to combine the benefits of memorization and generalization for recommender systems.
Guo et al.~\cite{guo2017deepfm} propose DeepFM that further integrates the power of factorization machines based on W\&D to learn high-order feature interactions for recommendations. 
Recently, advanced neural networks, such as attention-based models~\cite{song2019autoint}, have been developed. 
However, these techniques apply a fixed embedding dimension for all features, which would downgrade the model performance and consume substantial computational resources.

\subsection{AutoML for Recommendations}
\label{sec:bg_embedding_dim_search}
Automated Machine Learning (AutoML) has recently become a research hotspot due to its potential to automate the design process for recommender systems, minimizing human involvement. 
The research directions include feature selection~\cite{wang2022autofield, lin2022adafs}, embedding dimension search~\cite{zhaok2021autoemb, zhao2021autodim}, model architecture search~\cite{cheng2022towards}, and other component search~\cite{zhao2021autoloss, su2022detecting, su2021detecting}. 
Feature selection involves selecting a subset of field features in recommendation systems. 
For example, AutoField~\cite{wang2022autofield} uses a simple controller based on differentiable architecture search~\cite{liu2018darts} to select the top $K$ field features. 
AdaFS~\cite{lin2022adafs} enhances AutoField by modifying the controller to assign feature weights to fields for different samples.
The objective of embedding dimension search is to find mixed embedding sizes for each field.
For example, 
AutoEmb \cite{zhaok2021autoemb} finds the optimal dimension for each feature using differentiable search \cite{liu2018darts}. 
AutoDim \cite{zhao2021autodim} selects the best dimension for each field from a group of candidate dimensions in the same way as AutoEmb.
Model architecture search explores various network architectures and selects the optimal one~\cite{cheng2022towards}.

Our method is in alignment with embedding dimension search. 
All instances in the above methods share a uniform dimension size for each field. 
In contrast, our approach adaptively selects dimensions for each instance via the proposed Bernoulli gates, thereby considering the difference between individuals. 
Moreover, we introduce a polarization regularizer to overcome the shortcomings of the commonly-used top $K$ selection strategy.
Furthermore, rather than processing all samples through a single model, we propose to divide samples into clusters and train different recommender models with optimal dimensions tailored to different clusters. 
These unique and innovative designs in our proposed method have been proven effective in terms of both performance enhancement and inference cost saving.

\section{Method}

\begin{figure*}
    \centering
    \includegraphics[width=\textwidth]{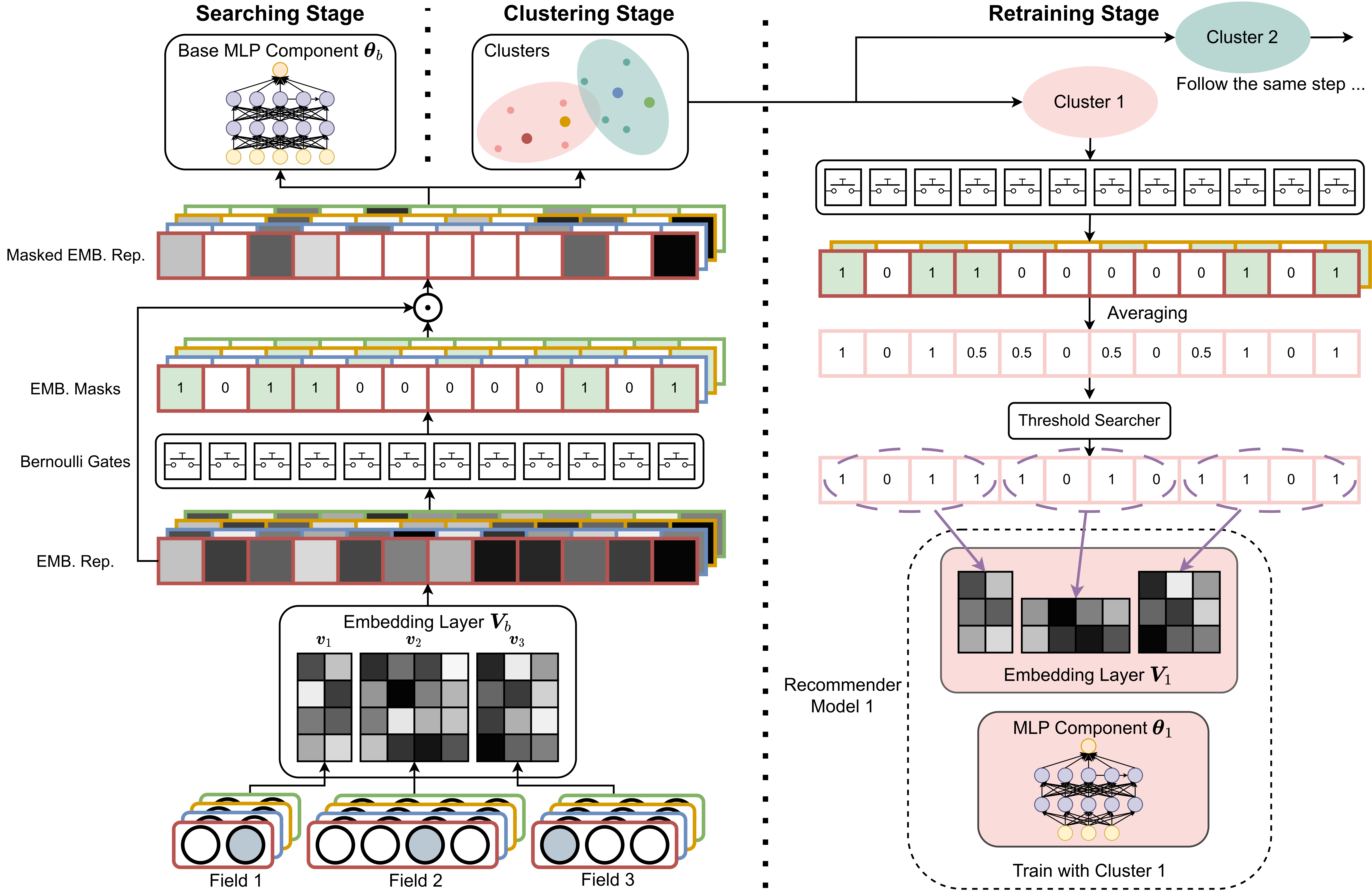}
    \caption{Overview of the three-stage training process in the iHAS framework. The four edge colors of the vectors (red, blue, yellow, and green) correspond to four different samples, which also correspond to the points in the clusters. The brightness level indicates the values of an element. ``$\bigodot$'' represents the element-wise product operation.}
    \label{fig:flowchart}
\end{figure*}

\begin{figure*}
    \centering
    \includegraphics[width=\textwidth]{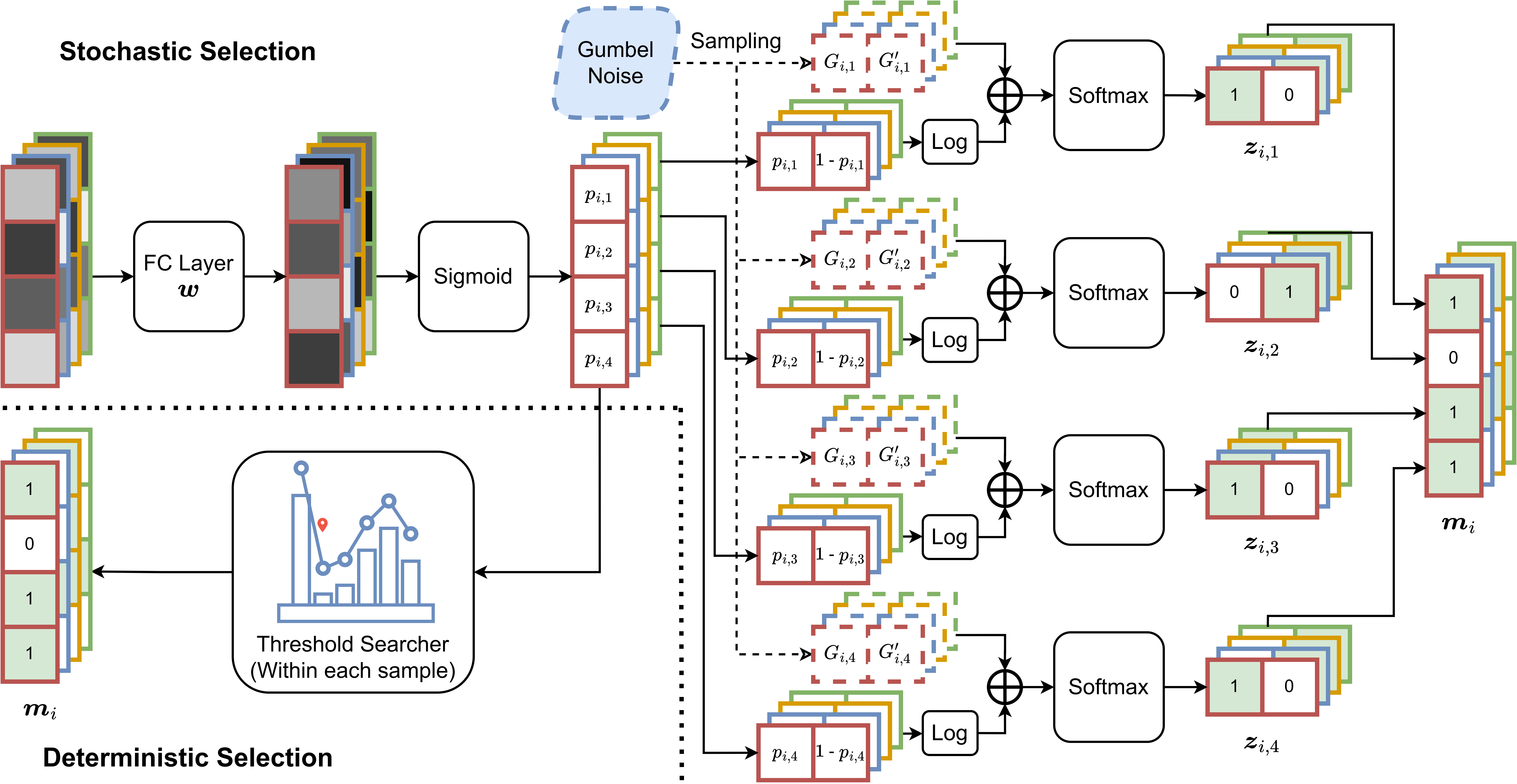}
    \caption{The detailed process of Bernoulli gates to generate embedding masks from the embedding representation. ``$\bigoplus$'' represents the element-wise summation operation.}
    \label{fig:bernoulligates}
\end{figure*}

This section introduces the technical specifications of the proposed
iHAS framework, as visualized in Figure~\ref{fig:flowchart}. 
We first provide a concise overview of the entire hierarchical training framework. Subsequently, the primary modules within our framework are described, as well as how to optimize them within each hierarchical stage.

\subsection{Overview}
The methodology for the iHAS framework comprises three stages: searching, clustering, and retraining.
It involves three principal modules: deep recommendation models (consisting of an embedding layer and an MLP component) for predicting user preferences, a Bernoulli gates layer responsible for dimension selection, and a K-means cluster algorithm that partitions the heterogeneous data.

In the searching stage, the key objective is to identify the optimal, instance-wise embedding dimensions across different fields, thus facilitating an accurate recommendation prediction. 
As shown in Figure~\ref{fig:flowchart}, the categorical field features are directed to an embedding layer to generate embedding representations. 
These representations are then processed through the Bernoulli gates to produce embedding masks using a stochastic selection mode (see Section~\ref{sec:stochastic_select}). 
Each embedding mask comprises a binary vector that serves as a gate on whether the corresponding dimension should be incorporated into the downstream architecture. 
Then the framework conducts an element-wise multiplication between a sample's embedding representations and embedding masks.
This resultant masked embedding representation is then directed to the base MLP component to predict user preference.

In the clustering stage, the main objective is to utilize K-Means algorithm~\cite{sculley2010web} to cluster samples into groups. 
This stage mostly mirrors the procedure used in the searching stage: 
using the embedding layer and Bernoulli gates (but using deterministic selection mode, see Section~\ref{sec:deterministic_select}) to calculate the masked embedding representations.  
These masked embedding representations are then used to train a mini-batch K-Means cluster~\cite{sculley2010web}. 
As shown in Figure~\ref{fig:flowchart}, the K-means separates the red and yellow samples from the blue and green samples.

The retraining stage aims to develop cluster-customized DLRMs, considering the variation in dimension patterns across different clusters.
In each cluster, we calculate the embedding masks (using deterministic selection mode) for all samples.
The resultant masks are then averaged to obtain one vector, which is used to determine the final embedding dimensions of each DLRM.

\subsection{Deep Learning Recommender Models}
In this subsection, we provide a brief introduction to the basic architecture of the DLRM. It typically comprises two primary components: an embedding layer and an MLP component.

\subsubsection{Embedding Layer}
In classic DLRM, the embedding layer is commonly used to convert 
the categorical inputs into a dense vector of real numbers. 

Let us denote the input of $N$ categorical field features for sample $i$ as $\boldsymbol{X}_i = [\Bfx{i, 1}, \cdots, \Bfx{i, n}, \cdots, \Bfx{i, N}]$, 
where $\Bfx{i, n} \in \mathbb{Z}^{|n|}$ represents the one-hot vector comprising sparse, high-dimensional binary values. The term $|n|$ denotes the number of unique values for $n$-th categorical field. 
For instance, a categorical field such as ``gender'' with unique values -- \texttt{male}, \texttt{female}, and \texttt{unknown} -- can be expressed through three-bit vectors $[1, 0, 0]$, $[0, 1, 0]$, and $[0, 0, 1]$, respectively. 
To process a numerical field feature, we will discretize it through custom-width binning, followed by applying a one-hot operation.
Then, the operation of the embedding layer can be represented as: 
\begin{equation*}
    \boldsymbol{e}_{i, n} = \boldsymbol{v}_{n} \, \Bfx{i, n},
\end{equation*} 
where $\boldsymbol{v}_{n} \in \mathbb{R}^{d \times |n|}$ is the embedding table of the $n$-th field, $d$ is the predefined embedding dimension (typically consistent across all fields), and $\boldsymbol{e}_{i, n}$ is the low-dimensional embedding representation. 
Therefore, the final embedding of the input data $\boldsymbol{X}_i$ through $N$ embedding tables is $\boldsymbol{E}_i = [\boldsymbol{e}_{i, 1}, \cdots, \boldsymbol{e}_{i, n}, \cdots, \boldsymbol{e}_{i, N}]$.

Notably, the embedding dimension search techniques we discussed earlier in Section~\ref{sec:bg_embedding_dim_search} (also the focus of this paper) aim at searching the optimal dimensions for embedding tables $\boldsymbol{V} = [\boldsymbol{v}_{1}, \cdots, \boldsymbol{v}_{n}, \cdots, \boldsymbol{v}_{N}]$.
Specifically, our goal is to discover the optimal individual embedding dimension for each field, given the inherent diversity in the heterogeneous dataset. 
This could potentially enhance prediction performance.

\subsubsection{MLP Component}

The MLP component plays a crucial role in DLRMs, tasked with encoding embedding representations and predicting the recommendation. Empirically, it comprises multiple fully-connected (FC) layers (characterized by parameter $\Bf{\theta}$) and is also equipped with non-linear activation functions such as ReLU~\cite{agarap2018deep} or Sigmoid, thereby facilitating the nonlinear encoding process of these representations.

In the iHAS system, we will train three different DLRMs, as illustrated in Figure~\ref{fig:flowchart}. Each DLRM consists of an embedding layer and an MLP component. These three models are named the base recommender model, recommender model 1, and recommender model 2, which are characterized by the parameter groups  $\{\Bf{V}, \Bf{\theta}\}_{b}$, $\{\Bf{V}, \Bf{\theta}\}_{1}$, and $\{\Bf{V}, \Bf{\theta}\}_{2}$, respectively.

\subsection{Bernoulli Gates}
\label{sec:bernoulligate}

Bernoulli gates operate as switches, facilitating the transmission of a sample's information from embedding tables to the downstream MLP component. 
Analogous to the $\mathcal{l}_0$ norm, we hope these ``switches'' to be capable of fully opening or closing without compromising the information integrity (shrinking the embedding representation). Inspired by the approach presented in~\cite{louizoslearning, yamada2020feature}, 
we use Bernoulli gates to predict each sample's relevant dimensions given its embedding representation. 
The detailed process of the Bernoulli gates is graphically depicted in Figure~\ref{fig:bernoulligates}.

The Bernoulli gates operate in two distinct modes: stochastic selection and deterministic selection. Under the stochastic selection mode, the gates operate as independent Bernoulli distributions, 
to independently ``open'' or ``close'' dimensions given the probabilities.  
The principle behind stochastic selection rests on the assumption that, given a sufficiently large number of training iterations, the gates will stochastically and comprehensively traverse all potential combinations of dimensions. 
This prompts the Bernoulli parameters to increase for beneficial dimensions and penalize unhelpful ones.

Once the Bernoulli distributions (gates) have been fully explored, we capitalize on the learned distribution by deterministically opening the most advantageous dimensions in the deterministic selection mode. However, learned Bernoulli probabilities often exhibit heavy-tailedness, making it challenging to distinguish between important and unimportant dimensions. To mitigate this, we suggest employing a polarization regularizer and an automatic threshold searcher (both discussed in Section~\ref{sec:polar}) inside Bernoulli gates.

\subsubsection{Stochastic Selection}
\label{sec:stochastic_select}
In our previous discussion, we aim for Bernoulli gates to function as independent Bernoulli distributions in the stochastic selection mode during the searching stage. 
The first objective is to encode the embedding representations to the desired independent Bernoulli probabilities. 
To this end, we employ an FC layer (with parameter $\Bf{w}$) and a Sigmoid activation layer ($\sigma$) to project these embedding representations of the $i$-th sample onto Bernoulli probabilities (upper left of Figure~\ref{fig:bernoulligates}), denoted by $\{p_{i, j}\}_{j=1}^{N^*} = \sigma (\Bf{w} \, \Bf{E}_i)$, where $N^* = N\times d$ is the total length of the embedding representations. 
This enables us to initiate a combinatorial search process over the space of Bernoulli probabilities and FC parameters.

However, optimizing a loss function, which includes discrete RVs (Bernoulli distributions), incurs high variance~\cite{mnih2016variational}. 
To overcome this obstacle, we adopt Gumbel-Softmax~\cite{jangcategorical} (aka Concrete distribution~\cite{maddisonconcrete}), which offers a viable continuous approximation to the Bernoulli distribution, as visualized in Figure 2 (right).

Recall that Gumbel-Max~\cite{gumbel1954statistical, maddison2014sampling} is an effective method for drawing samples from a Bernoulli distribution (or any type of discrete random variables), as long as we provide the class probabilities. 
For instance, if we use $p_{i, j}$ and $1 - p_{i, j}$ as the probabilities for selecting and not selecting the $j$-th embedding dimension for sample $i$, respectively, the Gumbel-Max trick to approximate Bernoulli distribution sampling can then be expressed as:
\begin{equation}
\label{eq:gumbel_max}
    \Bf{z}_{i, j} = \onehot \, (\argmax \, (\log p_{i, j} + G_{i, j} \ , \ \log (1 - p_{i, j}) + G'_{i, j})),
\end{equation}
where $G_{i, j}$ and $G'_{i, j}$ are i.i.d. samples drawn from a Gumbel distribution with the location at 0 and scale of 1, denoted as $\text{Gumbel} (0, 1)$.
However, both the ``$\onehot ()$'' and ``$\argmax ()$'' operations are non-differentiable, making them intractable for gradient descent optimization. 
Therefore, the softmax function is used as a continuous, differentiable approximation of these operations. 
The softmax function uses a temperature parameter $\tau \in \mathbb{R}^+$ to regulate the approximation degree (or the entropy of the distribution), as formalized:
\begin{equation}
\label{eq:gumbel_softmax}
    \Bf{z}_{i, j} = 
    \frac{
    \left[\exp (\, (\log p_{i, j} + G_{i, j}) \,/\, \tau \, ) \ , \ \exp (\,(\log (1-p_{i, j}) + G'_{i, j})\,/\, \tau \, ) \right]
    }{
    \exp (\,(\log p_{i, j} + G_{i, j})\,/\,\tau \,) \ + \ \exp (\,(\log (1 -p_{i, j}) + G'_{i, j})\,/\,\tau\,)} .
\end{equation}
As $\tau$ approaches 0, $\Bf{z}_{i, j}$ approximates the true binary vector, making the Gumbel-Softmax distribution become identical to the desired Bernoulli distribution. Then the final embedding masks, $\Bf{m}_i$, are created by concatenating the first bit of $\{\Bf{z}_{i, j}\}_{j=1}^{N^*}$.

However, our goal remains to produce true binary masks, which would effectively eliminate information from unimportant dimensions, as opposed to significantly shrinking them. 
The straight-through (ST) Gumbel-Softmax~\cite{jangcategorical, bengio2013estimating} serves well in this context. 
In the ST variant, the operation from Equation~\ref{eq:gumbel_max} is implemented in the forward pass while the continuous approximation from Equation~\ref{eq:gumbel_softmax} is used in the backward gradient descent. 
This approach enables sparse selection even when the temperature $\tau$ is high, while still allowing the gradient to propagate and update the parameters.

\subsubsection{Deterministic Selection}
\label{sec:deterministic_select}

\begin{figure}[t]
    \centering
    \vspace{-0.1in}
    \includegraphics[width=\columnwidth]{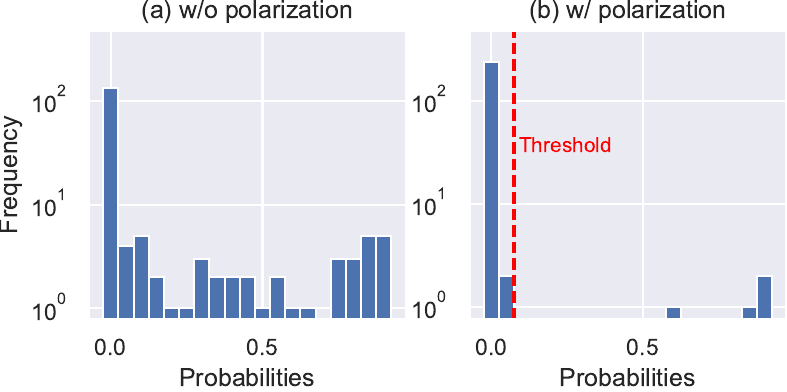}
    \vspace{-0.1in}
    \caption{Histogram of the Bernoulli probabilities for a sample from Avazu dataset, trained (a) with and (b) without polarization regularizer. Note that y-axes use log scales, and within the same range, to facilitate better visual comparison.}
    \label{fig:polarization}
\end{figure}

After training the Bernoulli probabilities ($p_{i,j}$) during the searching phase, we utilize these probabilities to determine which dimensions will contribute to the accuracy of the recommendation predictions. 
However, $p_{i,j}$ are characterized by high variance and heavy-tailedness, as shown by the histogram in Figure~\ref{fig:polarization} (left). 
These present two complications: 
(1) distinguishing important dimensions from unimportant ones becomes challenging; 
and 
(2) even the unimportant dimensions still possess a small probability of being selected. 
Moreover, masks created using Bernoulli gates introduce an element of randomness (Gumbel noise), which hinders their direct application during inference (where given the same data each time, consistent results should be generated).

To overcome these limitations, we propose a deterministic selection mode that directly selects the beneficial dimensions using the knowledge derived from the well-trained Bernoulli probabilities. 
This process is outlined in Figure~\ref{fig:bernoulligates} (bottom left). 
Firstly, we use the same FC layer and sigmoid layer to estimate the Bernoulli probabilities, $p_{i,j}$, analogous to the first step in the stochastic selection mode. 
Then, for each sample $i$, we search a threshold among the probabilities $\{p_{i, j}\}_{j=1}^{N^*}$ (see details in Section 3.3.3). 
We automatically adjust the gates to be open for probabilities exceeding this threshold and closed for those falling below it. 
The resulting embedding masks are utilized during the clustering and retraining phase (see Figure~\ref{fig:flowchart}).

\subsubsection{Polarization and Automatic Threshold Searcher}
\label{sec:polar}

Let's consider an empirical optimization procedure with an $\mathcal{l}_0$ regularization on the embedding masks during the searching stage:
\begin{align}
\label{eq:bce_loss}
    \mathcal{R} (\, \{\Bf{V}, \, \Bf{\theta}\}_b, \, \Bf{m} \,)   =  
    \mathbb{E}_{i} \, \mathbb{E}_{\Bf{m}_{i}} \left[ \mathcal{L}( \, f_{\Bf{\theta}_{b}}( \, \Bf{V}_b \cdot \Bf{X}_i \, \odot \, \Bf{m}_{i} \, ), \, y_i \,) \right], \\
\label{eq:l0_norm}
    \text{with regularizer} \quad \mathcal{R} (\Bf{m})  = \mathbb{E}_{i} \, \mathbb{E}_{\Bf{m}_{i}} \left[ \, \lambda \, \lVert \Bf{m}_{i}\rVert_0 \, \right], \quad \quad
\end{align}
where $\mathcal{L} (\cdot, \cdot)$ represents the binary cross-entropy (BCE) loss, $f_{\Bf{\theta}_{b}}(\cdot)$ represents the base MLP component with parameters $\Bf{\theta}_{b}$, $y_i$ is the ground truth label for the $i$-th sample, and $\lambda$ is a balancing factor for the regularizer. 
Since we use Gumbel-Softmax to produce the embedding masks, $\Bf{m}_{i}$, the term $\mathbb{E}_{i} \, \mathbb{E}_{\Bf{m}_{i}} \lVert \Bf{m}_{i}\rVert_0$ effectively equals to the sum of Bernoulli probabilities, $\sum_{i}\sum_{j=1}^{N^*} p_{i, j}$.

Our experimentation, however, indicates that simply using this $\mathcal{l}_0$ regularizer still experiences the heavy-tailedness and distinction difficulties, as discussed in Section~\ref{sec:deterministic_select}. 
Inspired from Zhuang et al.~\cite{zhuang2020neuron}, we incorporate a polarization regularizer into Equation~\ref{eq:l0_norm}:
\begin{align}
\label{eq:regularizer}
    \mathcal{R} (\Bf{m}) = \sum_{i} \sum_{j=1}^{N^*} \lambda \, p_{i, j} - \lvert  p_{i, j} -  \Bar{p}_{i}\rvert , \quad \text{with} \ \  \Bar{p}_{i} = \frac{1}{N^*} \sum_{j=1}^{N^*} p_{i, j} .
\end{align}
The intuition of the second term (which is the polarization regularizer) is to maximally distance $p_{i, j}$ from their mean $\Bar{p}_{i}$. Empirically, we have observed this polarization term effectively separates the probabilities of important and unimportant dimensions into two groups, thereby making them distinguishable (see Figure~\ref{fig:polarization}). 

Despite employing the regularizer as stated in Equation~\ref{eq:regularizer}, a threshold searcher is still required to identify the threshold for $p_{i,j}$. 
Currently, the histogram of $p_{i,j}$ trained with the polarization regularizer should have at least two peaks, with one of them located close to 0. 
Following the strategy from~\cite{zhuang2020neuron}, we scan the histogram from left to right to identify the first saddle point, i.e., the bin that contains the local minimum (the red pin in Figure~\ref{fig:bernoulligates}). 
The lower bound of this bin is subsequently set as the threshold.

\subsection{Clusters}
\label{sec:clusters}
Once obtaining a sample's unique embedding masks, our aim is to pass this sample through a recommender model tailored to its distinct dimension selection. 
This implies that, for every sample with its individual selected dimensions, there always exists a customized recommender model capable of handling this distinct input and generating a precise prediction.
Nonetheless, training an extensive number of recommender models is impractical due to the constraint of finite data size. 
For example, with a obviously underestimated dimension, $N^* = 100$, the number of possible combinations of dimensions is $\sum_{k=0}^{100} \binom{100}{k} = 2^{100}$, meaning at least $2^{100}$ distinct samples should be collected.

Consequently, we must strike a balance between using a one-size-fits-all recommender model (inadequate for highly heterogeneous samples) and using a fully-customized recommender model for each sample. 
Our resolution is to leverage clustering algorithms to segregate samples into separate groups. 
We anticipate that samples within the same group display similar patterns, which could be used to explore identical dimension patterns and identical recommender models. 
Furthermore, partitioning the data into clusters facilitates the training of small and manageable models for each cluster, thus accelerating the inference speed.

In this iHAS framework, we opt for the mini-batch K-Means clustering algorithm~\cite{sculley2010web} to partition the samples. 
We chose this variant due to its computational efficiency in handling large datasets (comprised of tens of millions of samples), and its capacity to avoid getting stuck at local optima.

\subsection{Optimization Strategy}

In the iHAS framework, we adopt a hierarchical training strategy to optimize different modules in different stages, inspired by~\cite{pham2018efficient, jethani2021have, wang2022autofield, lin2022adafs}, to tackle the encoding issue of jointly optimizing all modules.  
The detailed optimization strategy in each stage is illustrated in Algorithm~\ref{alg:optimization}.

\begin{algorithm}[t]
\caption{Optimization Strategy for iHAS}\label{alg:optimization}
\begin{algorithmic}[1]
\Require Training dataset $\Data_{\text{train}}$, validation dataset $\Data_{\text{val}}$ 
\Ensure Bernoulli gates parameters $\Bf{w}$, K-Means with two clusters, base recommender model $\{ \Bf{V}, \Bf{\theta}\}_b$, recommender model 1 $\{ \Bf{V}, \Bf{\theta}\}_1$, and recommender model 2 $\{ \Bf{V}, \Bf{\theta}\}_2$
\State \textbf{\#\#\# Searching \#\#\#}
\State Pretrain $\{\Bf{V}, \Bf{\theta}\}_b$ for 5 epoch using $\Data_{\text{train}}$
\While{not converge on $\Data_{\text{val}}$}
\State sample a mini-batch $\in \Data_{\text{train}}$, get $\Bf{m}$ by stochastic selection
\State update $\{\Bf{V}, \Bf{\theta}\}_b$ according the objective from Eq.~\ref{eq:bce_loss}
\State sample a mini-batch $\in \Data_{\text{val}}$, get $\Bf{m}$ by stochastic selection
\State update $\Bf{w}$ according the objective from Eq.~\ref{eq:bce_loss} and ~\ref{eq:regularizer}
\EndWhile

\State \textbf{\#\#\# Clustering \#\#\#} 
\While{not converge on cluster centroids}
\State sample a mini-batch $\in \Data_{\text{train}}$, get $\Bf{m}$ by deterministic selection, assign samples to their closest centroid
\State update the cluster centroids of K-means
\EndWhile

\State  \textbf{\#\#\# Retraining \#\#\#}
\State Based on the K-Means cluster, split $\Data_{\text{train}} \rightarrow \Data_{\text{train 1}}, \Data_{\text{train 2}}$, and split $\Data_{\text{val}} \rightarrow \Data_{\text{val 1}}, \Data_{\text{val 2}}$
\State find the optimal dimensions for $\Data_{\text{train 1}}$ via Bernoulli gates, and initialize $\{\Bf{V}, \Bf{\theta}\}_1$ with the corresponding dimensions
\While{not converge on $\Data_{\text{val 1}}$}
\State sample a mini-batch from $\Data_{\text{train 1}}$
\State update $\{\Bf{V}, \Bf{\theta}\}_1$ according the objective from Eq.~\ref{eq:bce_loss}
\EndWhile
\State repeat lines 16-20 with $\Data_{\text{train 2}}$, $\Data_{\text{val 2}}$, and $\{\Bf{V}, \Bf{\theta}\}_2$
\end{algorithmic}
\end{algorithm}

\subsubsection{Searching Stage}
As mentioned in~\cite{yamada2020feature}, the use of Gumbel-Softmax approximation for the discrete random variable suffers from high variance, which can lead to inconsistency in the set of selected dimensions. 
Inspired by~\cite{lin2022adafs}, we first pretrain the base recommender model for a few epochs to obtain a tentative reliable embedding representation. During these pretrain epochs, the Bernoulli gates are always deterministically open (no matter what embedding representation it receives). Then, after we've obtained a tentative reliable embedding representation, we initialize the parameters for the Bernoulli gates and start the stochastic selection. 
Later we adopt the bi-level optimization strategy~\cite{pham2018efficient, jethani2021have} to disjointly update the parameters $\Bf{w}$ in Bernoulli gates and the parameters $\{\Bf{V}, \Bf{\theta}\}_{\text{b}}$ in the base recommender model. 

\subsubsection{Clustering Stage}
In the clustering stage, we can obtain the masked embedding representations using the embedding tables and Bernoulli gates which have been trained during the searching stage. 
Remember we use deterministic selection mode for Bernoulli gates to generate embedding masks in this clustering stage. For each sample, compute its Euclidean distance to the centroid using the masked embedding representations, then assign it to the nearest centroid (group), and later update the centroids.

\subsubsection{Retraining Stage}
As to the retraining stage, we first divide all samples via the trained K-Means cluster. Then we find the majority dimensions for each group from their embedding masks using the deterministic mode.
After that, we initialize the deep recommender model 1 and 2 by their corresponding optimal dimensions and train them separately using the samples of each cluster.

\section{Experiment}
In this section, we conduct extensive experiments to evaluate our proposed framework. Specifically, the main research questions we care about are as follows:
\begin{itemize}
    \item \textbf{RQ1}: How does iHAS perform compared with other mainstream selection methods?
    \item \textbf{RQ2}: Can the proposed iHAS be successfully transferred to more powerful recommender models?
    \item \textbf{RQ3}: How does each component contribute to the overall performance of the proposed iHAS?
    \item \textbf{RQ4}: Does the proposed iHAS demonstrate efficiency when compared to baseline models?
    \item \textbf{RQ5}: Does iHAS construct rational recommender model structures?   
\end{itemize}
\subsection{Datasets}
\begin{table}[t]
  \caption{The statistics of Avazu and Criteo datasets.}
  \vspace{-0.1in}
  \label{tab:datasets}
  \begin{tabular}{cccccl}
    \toprule
    Dataset & \#Instances & \#Fields  & \#Features  \\
    \midrule
    Avazu   & 40,400,000       & 22       &  645,394  \\
    Criteo  & 45,840,617       & 39      & 1,086,810   \\
  \bottomrule
\end{tabular}
\end{table}
We conduct our experiments mainly on two commonly used public datasets, Avazu\footnote{\url{https://www.kaggle.com/c/avazu-ctr-prediction/}} and Criteo\footnote{\url{https://www.kaggle.com/c/criteo-display-ad-challenge/}}, which are both large-scale real-world datasets and serve as benchmarks in click-through rate (CTR) prediction tasks. 
Table ~\ref{tab:datasets} presents the detailed statistics of both datasets. Each dataset has been randomly segmented into training/validation/testing sets based on the proportions of 80\%, 10\%, and 10\%.
\begin{itemize}
    \item \textbf{Avazu} dataset consists of 40 million users' click records on ads over 11 days. Each record contains 22 categorical field features. Following the general preprocessing steps \cite{song2019autoint, zhu2021open}, we group fields of which frequency is less than ten as a single field ``others''.
    \item \textbf{Criteo} dataset consists of 46 million users' click records on display ads. Each record contains 26 categorical fields and 13 numerical fields. we use the preprocessing method as Avazu for the low-frequency fields (less than ten) and transform each numerical field $x$ by $log^2(x)$ if $x > 2$.
\end{itemize}

\begin{table*}[t]
\caption{Performance comparison between iHAS and baseline models.}
\vspace{-0.1in}
\label{tab:results_main}
\begin{tabular}{ccccccccc}
\toprule
Dataset   & Metric    & \multicolumn{7}{c}{Methods}   \\  \cmidrule(lr){3-9}
&                                              & PEP           & AutoField  & OptEmbed   & AdaFS-soft    & AdaFS-hard    & OptFS              & iHAS \\ \midrule
\multirow{2}*{Avazu} & AUC $\uparrow$          & 0.7665        & 0.7773     & 0.7630    & 0.7777        & 0.7763        & 0.7724                 & \textbf{0.7815}          \\
                     & Logloss $\downarrow$    & 0.3874        & 0.3813     & 0.3894     & 0.3812        & 0.3821        & 0.3840                &  \textbf{0.3791}         \\ \midrule
\multirow{2}*{Criteo}   & AUC $\uparrow$       & 0.8006        & 0.8029       &  0.7962  & 0.8039        & 0.8031        & 0.8015                &\textbf{0.8043}           \\
                        & Logloss $\downarrow$ & 0.4507        & 0.4490    & 0.4543     & 0.4484        & 0.4560        & 0.4504                 &\textbf{0.4478}           \\ \bottomrule  
\end{tabular}
\end{table*}

\subsection{Evaluation Metrics}
Following the previous works \cite{qu2022single, liulearnable}, 
we evaluate the performance of our method using two common metrics: \textbf{AUC} and \textbf{Logloss}. 
AUC refers to the area under the ROC curve, which means the probability that a model will rank a randomly selected positive instance higher than a randomly selected negative one.
A higher AUC value indicates superior model performance.
On the other hand, Logloss, aka binary cross-entropy loss, directly quantifies the model's performance, with a lower score denoting more accurate predictions.
Note that a marginal \textbf{0.001-level} improvement in AUC (increase) or Logloss (decrease) is perceived as a significant enhancement in model performance \cite{zhao2021autodim, wang2022autofield, lyu2023optimizing}.

\subsection{Baseline Methods}
We compare our proposed method with the following state-of-the-art methods:
\begin{itemize}
    \item  \textbf{PEP} \cite{liulearnable}: It adopts trainable thresholds to prune redundant embedding dimensions.
    \item \textbf{AutoField} \cite{wang2022autofield}: It utilizes neural architecture search techniques \cite{liu2018darts} to select important field features.
    \item \textbf{OptEmbed} \cite{lyu2022optembed}: It trains a
    supernet with various selected embedding dimensions, then uses evolution search to find the optimal embedding dimensions based on the supernet.
    \item \textbf{AdaFS} \cite{lin2022adafs}: It assigns weights to different fields in a soft manner (AdaFS-soft) or masks unimportant fields in a hard manner (AdaFS-hard) via a novel controller network.
    \item \textbf{OptFS} \cite{lyu2023optimizing}: It simultaneously selects optimal field features and the optimal interactions between these features using ``binary gates''.   
\end{itemize}

\subsection{Implementation Details}
We implement our method based on a public library\footnote{\url{https://github.com/rixwew/pytorch-fm}} that involves sixteen commonly-used DLRMs. 
As our framework is model-agnostic, it can be seamlessly integrated with any of these models, see Section~\ref{sec:transfer}. 
For the embedding layer, we set the initial embedding size of all fields as 16 in accordance with the previous works \cite{lin2022adafs, wang2022autofield}. 
For the MLP component, we adopt two fully-connected layers of size $(16, 8)$ with the ReLU activation function. 
We use Adam optimizer \cite{kingma2014adam} with an initial learning rate of $0.001$, and weight decay of 1e-6.
The batch size is set to 2048. 
We sample one validation batch every 100 training batches for bi-level optimization. 
The temperature $\tau$ for ST Gumbel-Softmax is set to $0.1$. 

The baseline models are implemented by the codes provided by their authors. For a fair comparison, we set the initial embedding dimension as 16 for all baselines. All the experiments are run on a single machine with an Nvidia RTX 3090 GPU.

\subsection{Overall Performance (RQ1)}
Table \ref{tab:results_main} compares the overall performance of our proposed iHAS and other baseline models on the Avazu and Criteo datasets.
We summarize our observations below.

First, our iHAS outperforms all the state-of-the-art baseline methods as it can achieve higher AUC and lower Logloss on both datasets, demonstrating the effectiveness of iHAS in deep recommendation systems. 
Specifically, iHAS outperforms the runner-ups by $0.0038$ (AUC) and $0.0021$ (Logloss) on the Avazu datasets, and by $0.0004$ (AUC) and $0.0006$ (Logloss) on the Criteo datasets. 

Secondly, among all baselines, AdaFS-soft is the most effective model for the Avazu and Criteo datasets. 
However, it only shrinks the field features using a feature weighting layer and therefore does not completely eliminate the effect of unimportant fields.  
Although AdaFS-hard attempts to mask unimportant fields by uniformly keeping the top $K$ features,
the trained feature weights may still exhibit a high variance pattern (remember the non-distinguishable probabilities in Figure~\ref{fig:polarization}, left panel).
Therefore, this top $K$ selection manner may lead to selecting unimportant features or omitting the important feature in their final model, further compromising the model performance. Our polarization regularizer and threshold searcher can help with this issue, as detailed in Section~\ref{sec:polar} and evidenced by the empirical ablation study in Section~\ref{sec:ablation}.

Lastly, other baselines apply global feature/dimension selection across all samples, which fails to account for inherent variations among heterogeneous individuals, and consequently leads to suboptimal performance.
Additionally, PEP mainly emphasizes on the model size, i.e., it stops searching once the embedding table reaches a predefined parameter size.
This approach may result in a sub-optimal embedding table due to overlooking the model performance. 
AutoField\footnote{The performance score for AutoField is borrowed from its original paper~\cite{wang2022autofield} as they use the same experimental settings and have not publicly released the codes.} 
also employs the top $K$ selection manner, again leading to feature misselection and inferior model performance.

\subsection{Transferability Analysis (RQ2)}
\label{sec:transfer}
\begin{table}[t]
\caption{Transferability of iHAS on the Avazu dataset.}
\vspace{-0.1in}
\label{tab:transferability}
\begin{tabular}{ccccc}
\toprule
Model                 & Metric                & \multicolumn{3}{c}{Transfer Type}                                                                                     \\  \cmidrule(lr){3-5}
                        &     & Original    & AdaFS-soft        & iHAS      \\ \midrule
\multirow{2}*{FM}    & AUC $\uparrow$        & 0.7766                  & 0.7799        &\textbf{0.7826}         \\
                     & Logloss $\downarrow$  & 0.3815                   &0.3797        &\textbf{0.3793}        \\ \midrule
\multirow{2}*{W\&D}   & AUC $\uparrow$        & 0.7772                  &0.7790         &\textbf{0.7797}   \\
                        & Logloss $\downarrow$  & 0.3815                 & 0.3802       &\textbf{0.3800}  \\ \midrule  
\multirow{2}*{DeepFM}   & AUC $\uparrow$        & 0.7806               & 0.7817       &\textbf{0.7840}  \\
                        & Logloss $\downarrow$  & 0.3795                   & 0.3786       &\textbf{0.3784}     \\ \bottomrule 
\end{tabular}
\end{table}
In this subsection, we explore the transferability of iHAS. Specifically, we freeze the parameters of the well-trained Bernoulli gates and utilize them to help train other popular deep recommendation models, including FM~\cite{rendle2010factorization}, W\&D~\cite{cheng2016wide}, and DeepFM~\cite{guo2017deepfm}. 

Table~\ref{tab:transferability} shows the experimental results on Avazu, where ``original'' refers to the corresponding model without any selection. 
We can observe that: (i) all the recommendation models have great improvement by adopting iHAS, which again demonstrates the importance of performing selection in the recommendations; (ii) The transferability of iHAS is better than the best baseline (by comparing the iHAS and AdaFS-soft in Table~\ref{tab:transferability}), which validates the effectiveness of our Bernoulli gates.

In summary, we conclude that iHAS has outstanding transferability across different recommendation models, which enables it to be leveraged in complicated real-world recommender systems.

\subsection{Ablation Study (RQ3)}
\label{sec:ablation}
\begin{table}[t]
\caption{Ablation study on the Avazu datasets.}
\vspace{-0.1in}
\label{tab:ablation}
\setlength\tabcolsep{3.7pt}
\begin{tabular}{ccccccc}
\toprule
                  \multirow{2}*{Metric}                & \multicolumn{5}{c}{Methods}                                                                                     \\  \cmidrule(lr){2-7}
                             &Base & iHAS-1    & iHAS-2     & iHAS-3  & iHAS-4 & iHAS      \\ \midrule
     AUC $\uparrow$        &0.7765 &0.7772          &0.7767         &0.7801       &0.7768   & \textbf{0.7815}        \\
    Logloss $\downarrow$  &0.3818 &0.3813        &0.3816         &0.3800      &0.3817    &\textbf{0.3791}        \\ \bottomrule
\end{tabular}
\end{table}
In this subsection, we conduct the ablation study of key components in iHAS, as shown in Table~\ref{tab:ablation}. 
The Base model keeps all fields and the uniform embedding dimensions without any selections, and
we derive four variants from iHAS: 
(i) iHAS-1: This variant is the model directly obtained in the searching stage, i.e., we remove the clustering and retraining stages;
(ii) iHAS-2: This variant consists of a searching stage and a retraining stage. After we have the well-trained Bernoulli gates, we select dimensions across all samples to retrain one recommender model instead of separating samples into different clusters for retraining cluster-customized recommender models;
(iii) iHAS-3: This variant is the standard iHAS without using the polarization regularizer described in Section \ref{sec:polar};
(iv) iHAS-4: This variant doesn't consider instance-wise differences by disconnecting the Bernoulli probabilities with the embedding representations of each sample. That means the Bernoulli probabilities become $\{p_{j}\}_{j=1}^{N^*} = \sigma(w^*)$ where $w^*$ is simply a randomly initialized vector of the dimension length, making every sample share the same probability for every dimension. 

Based on the results in Table \ref{tab:ablation}, we can find: 
(i) iHAS and its variants can increase the AUC and decrease the Logloss compared with the Base model, which indicates the necessity of performing selection on embedding dimensions for boosting model performance;
(ii) iHAS performs better than iHAS-1, which indicates the necessity of the subsequent clustering and retraining stages;
(iii) iHAS-1 outperforms iHAS-2, therefore, separating instances into different groups, i.e., the clustering stage, is beneficial for boosting the performance;
(iv) Polarization is vital for acquiring better Bernoulli gates by comparing iHAS with iHAS-3;
(v) Respecting the difference between instances can further boost the performance by comparing iHAS with iHAS-4. 

\subsection{Efficiency Analysis (RQ4)}
\begin{figure}
    \centering
    \vspace{-0.1in}
    \includegraphics[width=\columnwidth]{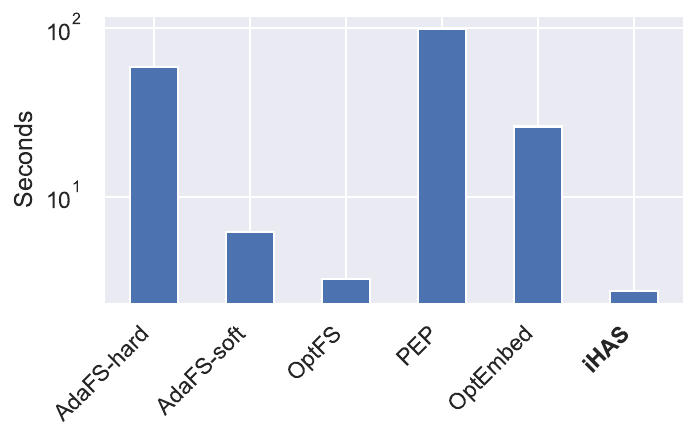}
    \vspace{-0.4in}
    \caption{Inference time (in log scale) of iHAS and other baselines on the Avazu dataset.}
    \label{fig:inference_time}
\end{figure}
In addition to model performance, efficiency is vital when deploying the recommendation model into online systems, especially inference efficiency. We report the inference time on the whole test set of iHAS and other baselines in Figure \ref{fig:inference_time}.
We can find that iHAS achieves the least inference time. This is because iHAS feed different test data into its preferred recommender model of smaller size instead of feeding all test data into a single model which may lead to additional inference cost on some data.
On the contrary, PEP requires the longest inference time because its embedding table is usually sparse and hardware-unfriendly.

\subsection{Case Study (RQ5)}

\begin{figure}
    \centering
    \vspace{-0.1in}
    \includegraphics[width=\columnwidth]{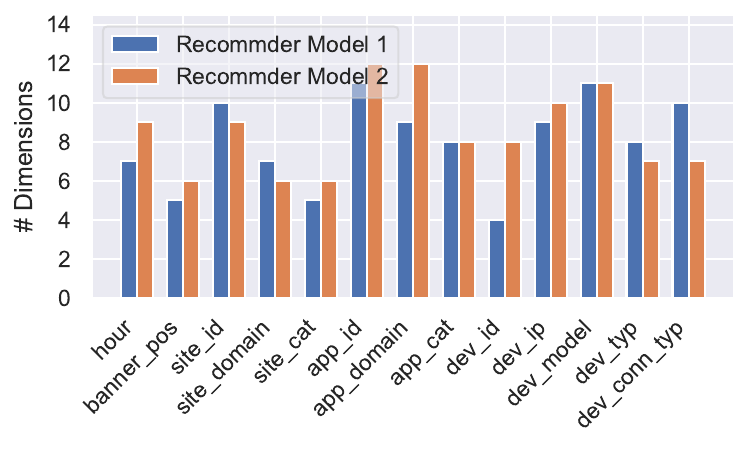}
    \vspace{-0.3in}
    \caption{Case study of selected dimensions of each field for each DLRM in iHAS on the Avazu dataset.}
    \label{fig:dimension_size}
\end{figure}

In this subsection, we first use a case study to investigate the optimal embedding dimensions for each cluster from iHAS. We show the results on Avazu as an example and exclude all anonymous field features in Figure \ref{fig:dimension_size}.

\begin{figure}
    \centering
    \vspace{-0.1in}
    \includegraphics[width=0.8\columnwidth]{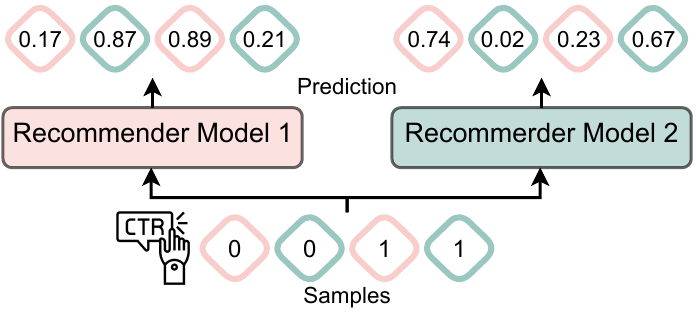}
    \vspace{-0.1in}
    \caption{Example data predictions on the Avazu dataset, where the samples with the same edge color belong to the same cluster. The ground truths and prediction scores are displayed in the diamonds.}
    \label{fig:benefit_of_cluster}
\end{figure}

We can observe that:
(i) Each field's optimal dimensions greatly vary from one to another (from 4 to 12), which highlights the necessity of dimension search in recommender systems;
(ii) id-related features, e.g., site\_id and app\_id, typically possess more dimensions. This aligns with human intuition as the id-related features are the core of recommender systems;
(iii) Samples within different clusters tend to select different dimensions for each field, which validates our claim that different clusters present different patterns and should be trained separately to enhance performance and reduce inference time in Section \ref{sec:clusters}. 

Furthermore, we use four samples to illustrate the effectiveness of the iHAS framework consisting of group-customized recommender models.
Figure~\ref{fig:benefit_of_cluster} shows four samples grouped into two clusters (two in pink and two in cyan). 
Each cluster has its customized recommender model.
We can find that the predictions are more correct (lower Logloss) if we feed the sample into its corresponding model. However, if feeding all of them together into one of the recommender models, we will receive some wrong predictions. 

\section{Conclusion}
This study proposes an instance-wise Hierarchical Architecture Search framework, iHAS, as an innovative solution to the challenges associated with identifying optimal embedding dimensions for DLRMs. 
iHAS employs a three-stage hierarchical training strategy including searching, clustering, and retraining.
The searching stage aims to identify the optimal embedding dimensions for each sample across different fields. 
Subsequent stages of clustering and retraining provide a mechanism for gathering similar samples as clusters and training cluster-customized DLRMs based on the individual optimal dimensions, thereby enhancing recommendation predictions.
We conduct extensive experiments on two large-scale datasets to authenticate the efficacy of the proposed framework. 
The results demonstrate that iHAS could boost the performance of deep recommendations while reducing inference costs. 
Additionally, iHAS exhibits outstanding transferability to popular DLRMs.

\bibliographystyle{ACM-Reference-Format}
\balance
\bibliography{sample-base}




\end{document}